\documentstyle[12pt]{article}
\def\1{\mbox{I\hspace{-.15em}1}}

\def\b{\begin{equation}}
\def\e{\end{equation}}
\def\bee{\begin{enumerate}}
\def\eee{\end{enumerate}}
\title{A Natural Renormalization of\\ the One-Loop Effective Action\\ for Scalar Field
in Curved Space-time }
\author{ Mohammad Vahid Takook \thanks{e-mail: takook@razi.ac.ir}}
\date{\today}

\begin{document}
\maketitle {\it \centerline{  Department of Physics, Razi
University, Kermanshah, IRAN}}

\begin{abstract}

It has been shown that the negative norm states necessarily appear
in a covariant quantization of the free minimally coupled scalar
field in de Sitter space \cite{dere,gareta1}. In this process
ultraviolet and infrared divergences have been automatically
eliminated \cite{ta3}. A natural renormalization of the one-loop
interacting quantum field in Minkowski space-time ($\lambda\phi^4$
theory) has been achieved through the consideration of the
negative norm states \cite{ta4}. One-loop effective action for
scalar field in a general curved space-time has been calculated by
this method and a natural renormalization procedure in the
one-loop approximation has been established.

\end{abstract}

\vspace{0.5cm} {\it Proposed PACS numbers}: 04.62.+v, 03.70+k,
11.10.Cd, 98.80.H \vspace{0.5cm}
\newpage
\section{Introduction}

In the previous papers, we have shown the necessity of keeping the
negative norm states, for a fully covariant quantization of the
minimally coupled scalar field in de Sitter space (``Krein'' QFT)
\cite{dere,gareta1,ta3}. We have also shown that the effect of
these unphysical states appears in the physics of the problem by
allowing an automatic renormalization of the theory in the
one-loop approximation. We have also shown that, for the physical
states (positive norm states), the energy is positive whereas for
the negative norm states (called ``unphysical'' states) the energy
is negative.

Consideration of the negative norm states was proposed by Dirac
 in 1942 \cite{di}. In 1950 Gupta applies the idea in QED \cite{gu}. The presence of higher derivative in
the Lagrangian also lead to ghosts, states with negative norm
\cite{hato}. Mathematically for preserving the covariant
principle, the auxiliary negative norm states were presented.
Their presence has also different consequences for example in QED
the negative energy photon disappear \cite{gu}, and in de Sitter
the infrared divergence eliminated \cite{gareta1}. The physical
interpretation however is not yet clear and any further
interpretation needs far more investigations
\cite{rami,he,viba,bl}.

In the usual QFT, the one-loop effective action for the scalar
field in a general curved space is divergent \cite{bida}. One way
to remove this divergence is modifying the Einstein's field
equations. Precisely for this reason that one of the most
important problems of  quantum gravity appears: for every new loop
expansion,  new terms in the Einstein's field equations are
needed. This means that the theory is not renormalizable.  In this
paper we have proposed a method of field quantization for
calculating of the one-loop effective action for the scalar field
in a general curved space-time, which result in a finite one-loop
effective action.

\section{Usual QFT calculation}

Recalling the usual QFT calculation for the one-loop effective
action \cite{bida}, the scalar field in a general curved
space-time is defined by: \b [\Box +m^2+\xi R(x)]\phi(x)=0,\e
where $\Box$ is the Laplace-Beltrami operator in curved space-time
and $m$ is the ``mass'' of the field quanta. $R(x)$ is the Ricci
scalar curvature and $\xi$ is the coupling constant between the
scalar field and the gravitational field. The adiabatic expansion
of the Feynman propagator is defined by \cite{bida} \b
G_F^p(x,x')\approx (-g(x))^{\frac{-1}{4}}\int
\frac{d^4k}{(2\pi)^4}e^{-ik.y}\left[\sum_{j=0}^\infty
a_j(x,x')(-\frac{\partial}{\partial
m^2})^j\right]\frac{1}{k^2-m^2+i\epsilon}, \e where symbol
$\approx$ indicates an asymptotic expansion and $y$ is the Reimann
normal coordinates for the point $x$, with origin at point $x'$.
In the semi classical theory the Einstein's field equations are:
\b R_{\mu \nu}-\frac{1}{2}R g_{\mu \nu}+\Lambda g_{\mu \nu}=-8\pi
G <T_{\mu \nu}>,\e where $<T_{\mu \nu}>$ is the quantum
expectation value of the matter stress-tensor.  The effective
action ($W$) of the quantum matter field is defined by : \b
\frac{2}{(-g)^{\frac{1}{2}}} \frac{\delta W }{\delta g^{\mu
\nu}}=<T_{\mu \nu}>.\e In the one-loop approximation it is defined
by \b W= \int d^4x[-g(x)]^{\frac{1}{2}}
L_{eff}^p(x)=-\frac{1}{2}i\int d^4x[-g(x)]^{\frac{1}{2}}<x\mid \ln
(- G_F^p)\mid x>,\e where \b  <x\mid G_F^p\mid x'>=G_F^p(x,x').\e
We can also write the effective Lagrangian in the following form
\b L_{eff}^p(x)=-\frac{1}{2}i<x\mid \ln (- G_F^p)\mid
x>=\frac{1}{2}i\int_{m^2}^\infty dm^2G_F^p(x,x).\e Replacing
$G_F^p(x,x)$ from equation $(2)$ in equation $(7)$ we have \b
L_{eff}^p(x)\approx
\frac{i}{2}(-g(x))^{\frac{-1}{4}}\int_{m^2}^{\infty} dm^2 \int
\frac{d^4k}{(2\pi)^4}\left[\sum_{j=0}^\infty
a_j(x)(-\frac{\partial}{\partial
m^2})^j\right]\frac{1}{k^2-m^2+i\epsilon}.\e In this relation
there are three terms which diverge. For eliminating these
divergences the left-hand side of the Einstein's field equation is
changed and $\Lambda$ and $G$ are also redefined (renormalization
procedure) \cite{bida} \b R_{\mu \nu}-\frac{1}{2}R g_{\mu
\nu}+\Lambda g_{\mu \nu}+a ^{(1)}H_{\mu \nu}+b ^{(2)} H_{\mu
\nu}.\e A standard technique in QFT indicates that the terms
involving higher derivatives of the metric are expected in view of
divergence elimination.

Using following integral representation in $(2)$ $$
\frac{1}{k^2-m^2+i\epsilon}=-i\int_0^\infty ds
e^{is(k^2-m^2+i\epsilon)},$$ permuting the $d^4k$ integration with
$ds$ integration in $(2)$, and performing the former, the Green's
function can be written in terms of the Bessel functions
\cite{bida} $$ G_F^p(x,x') \approx
\Delta^{\frac{1}{2}}(x,x')\left[\sum_{j=0}^\infty
a_j(x,x')(-\frac{\partial}{\partial m^2})^j\right]
G_F^{p(M)}(x,x')$$ where $\Delta$ is the Van Vleck determinant and
$G_F^{p(M)}(x,x')$ is the Feynman Green function in the Minkowski
space $$ G_F^{p(M)}(x,x')=-\frac{1}{8\pi}\delta
(\sigma)+\frac{m^2}{8\pi}\theta(\sigma)\frac{J_1
(\sqrt{2m^2\sigma})-iN_1 (\sqrt{2m^2\sigma})}{\sqrt{2m^2
\sigma}}$$ $$ -\frac{im^2}{4\pi^2}\theta(-\sigma)\frac{K_1
(\sqrt{-2m^2\sigma})}{\sqrt{-2m^2
\sigma}},\;\;\;\sigma=\frac{1}{2}(x-x')^2.$$

\section{Krein QFT calculation}

The origin of divergence lies in the singular character of Green's
function at short relative distances. It has been shown in
\cite{dere,gareta1,ta3,ta4} that if the unphysical negative norm
states are taken into account in the field quantization, the
time-ordered product of fields or the ``Feynman'' propagator  \b
iG_T(x,x')=<0\mid T\phi(x)\phi(x') \mid 0>,\e is defined by the
following relation \b
G_T(x,x')=\frac{1}{2}[G_F^p(x,x')+(G_F^{p}(x,x'))^*],\e where
$G_F^p(x,x')$ is usual Feynman propagator. Similar to the
quantization of the electromagnetic field in Minkowski space
\cite{itzu}, insofar as only average values are observed, we see
that the unphysical negative norm states disappeared when
restricting ourselves to physical states although the Green's
function is changed due to presence of negative norm states.

In this method the time-ordered product two-point function is
$$ G_T(x,x')\approx
\Delta^{\frac{1}{2}}(x,x')\left[\sum_{j=0}^\infty
a_j(x,x')(-\frac{\partial}{\partial
m^2})^j\right]\left[\frac{m^2}{8\pi}\theta(\sigma)\frac{J_1
(\sqrt{2m^2\sigma})}{\sqrt{2m^2 \sigma}}-\frac{1}{8\pi}\delta
(\sigma)\right]$$ \b
=\Delta^{\frac{1}{2}}(x,x')\left(-\frac{a_0}{8\pi}\delta
(\sigma)+\left[\sum_{j=0}^\infty
a_j(x,x')(-\frac{\partial}{\partial
m^2})^j\right]\frac{m^2}{8\pi}\theta(\sigma)\frac{J_1
(\sqrt{2m^2\sigma})}{\sqrt{2m^2 \sigma}}\right).\e This expression
in the limit $x\rightarrow x'$ and $\sigma>0$, simplifies to \b
\lim_{x\rightarrow x'}G_T(x,x') \approx \frac{1}{16
\pi}[a_0(x)m^2-a_1(x)]\int_0^\infty se^{-s}ds,\e where the
integral $\int_0^\infty se^{-s}ds=1$ is presented in view of the
following calculation of the effective action. The divergence of
the delta function form $(\delta(\sigma=0))$, is ignored. This
term produce a constant term in the effective action.

If we use eq. $(13)$ and the procedure which is used when only
positive norm states are involved \cite{bida}, the following
expression for the effective Lagrangian is obtained \b
L_{eff}(x)=-\frac{i}{2}\lim_{x\rightarrow x'}<x \mid \ln (-
G_T)\mid x'>=-\frac{i}{2}\lim_{x\rightarrow x'}\int_{0}^\infty
<x\mid e^{-iKs}\mid x'> (is)^{-1}ids,\e where \b
\lim_{x\rightarrow x'} <x\mid e^{-iKs}\mid x'>\approx
\frac{i}{16\pi}[a_0(x)m^2-a_1(x)]m^4se^{-m^2 s}.\e Then the
effective Lagrangian in the one-loop approximation reads: \b
L_{eff}(x)\approx \frac{1}{32 \pi}[a_0(x)m^2-a_1(x)]m^4
\int_0^\infty e^{-m^2 s} ds =
\frac{1}{32\pi}[a_0(x)m^2-a_1(x)]m^2.\e By using the following
relations \cite{bida}: \b a_0(x)=1\;\;,\;\;\;\; a_1(x)=
(\frac{1}{6} -\xi)R(x),\e  the effective action give \b
L_{eff}(x)\approx
 \frac{m^4}{32\pi}-(\frac{1}{6}
-\xi)\frac{m^2}{32\pi}R(x).\e One of the interesting issue of this
calculation is that, in the one-loop approximation, the effective
action is similar to the Einstein-Hilbert action with a
cosmological constant (de Sitter background).

\section{Conclusion}

't Hooft and Veltman have shown that all one-loop divergencies of
pure  gravity can be absorbed in a field renormalization. However,
in the case of interaction of gravity and scalar field,
divergencies in physical quantities are not eliminated
\cite{thve}, unless new terms are introduced to the Einstein's
field equation \cite{bida}. In this paper, it is proved that if
the unphysical negative norm states are considered in QFT
formalism, non-divergent one loop-approximation of the interaction
of gravity with a scalar field can be archived {\it i.e.} the
effective action for scalar field is naturally convergent. Thus
the Einstein's field equation is not altered. The quantum gravity
in the one-loop approximation behaves in a non anomalous way.
\vskip 0.5 cm

\noindent {\bf{Acknowlegements}}: The author would like to thank
S. Rouhani for very useful discussions and  D.G. Chakalov for
their interest in this work. I would like to thank the referee for
very useful suggestions.


\begin{thebibliography}{a}

%\addcontentsline{toc}{chapter}{Bibliographie}

\bibitem{dere} De Bi\`evre S., Renaud J., Phys. Rev. D,
$57(1998)6230$
\bibitem{gareta1}  Gazeau J. P., Renaud J., Takook M.V., Class. Quantum
Grav., $17(2000)1415$, gr-qc/$9904023$
\bibitem{ta3} Takook M.V., Mod. Phys. Letters A, $16(2001)1691,
gr-qc/0005020$
\bibitem{ta4} Takook M.V., Int. J. Mod. Phys. E, $11(2002)509,
gr-qc/0006019$
\bibitem{di} Dirac P.A.M., Proc. Roy. Soc. A, $180(1942)1$
\bibitem{gu} Gupta S.N., Proc. Phys. Soc. Sect. A, $63(1950)681$
\bibitem{hato} Hawking S.W., Hertog T., Phys. Rev. D, $65 (2002) 103515$
\bibitem{rami} Ramirez A., Mielinik B., Rev. Mex. Fis.,
$49S2(20003) 130$
\bibitem{he} Helfer A.D., {\it The Physics of Negative Energy Densities}, hep-th/9811081
\bibitem{viba} Visser M., Barcelo C., {\it Energy condition and their cosmological implications}, gr-qc/0001099
\bibitem{bl} Blinnikov S., Surveys High Energ. Phys. 15(2000)37, astro-ph/9911138
\bibitem{bida} Birrell N.D., Davies P.C.W., Cambridge University Press,  $(1982)$
{\it QUANTUM FIELD IN CURVED SPACE}
\bibitem{itzu} Itzykson C., Zuber J-B., McGraw-Hill, Inc. $(1988)$  {\it Quantum Field Theory}
\bibitem{thve} 't Hooft G., Veltman M.,  Ann. Inst. Henri Poincar\'e, Vol XX, $n^0$ 1, $(1974)$69

\end{thebibliography}
\end{document}